\documentclass[journal]{IEEEtran}

\usepackage{pgfplots}
\usepackage[bookmarks,colorlinks]{hyperref}
\usepackage[linesnumbered,ruled,lined]{algorithm2e}
\usepackage{enumitem}
\usepackage{algpseudocode}
\usetikzlibrary{shapes.multipart,intersections}
\usepackage{cite}
\usepackage{amsmath,amssymb,amsfonts,amsthm,steinmetz}
\usepackage{graphicx}
\usepackage{mathrsfs}  
\usepackage{textcomp}
\usepackage{acronym}
\usepackage{xcolor}
\usepackage{upgreek,xspace}
\usepackage{array}
\usepackage{tikz}
\usetikzlibrary{calc}
\makeatletter
\newcommand{\gettikzxy}[3]{%
  \tikz@scan@one@point\pgfutil@firstofone#1\relax
  \edef#2{\the\pgf@x}%
  \edef#3{\the\pgf@y}%
}
\makeatother

\usepackage[draft]{todonotes}

\usepackage{esvect}

\usepackage{times}
\usepackage{bm}
\usepackage{amsmath}
\usepackage{amssymb}
\usepackage{stmaryrd}
\usepackage{babel}
\usepackage{graphics, graphicx}
\usepackage{xcolor}
\usepackage{gensymb}
\usepackage{cite}
\usepackage{enumitem}
\usepackage{url}

\begin{document}

\title{Scalable Multiport Antenna Array Characterization \\with PCB-Realized Tunable Load Network Providing Additional ``Virtual'' VNA Ports}

\author{Jean~Tapie~and~Philipp~del~Hougne,~\IEEEmembership{Member,~IEEE}
\thanks{This work was supported in part by the ANR France 2030 program (project ANR-22-PEFT-0005), the ANR PRCI program (project ANR-22-CE93-0010), the French Defense Innovation Agency (project 2024600), the French Region of Brittany (project TUNSY), the European Union's European Regional Development Fund, and the French Region of Brittany and Rennes Métropole through the contrats de plan État-Région program (projects ``SOPHIE/STIC \& Ondes'' and ``CyMoCoD'').}
\thanks{
J.~Tapie and P.~del~Hougne are with Univ Rennes, CNRS, IETR - UMR 6164, F-35000, Rennes, France (e-mail: \{jean.tapie; philipp.del-hougne\}@univ-rennes.fr).
}
\thanks{\textit{(Corresponding Author: Philipp del Hougne.)}}
}

\maketitle

\begin{abstract}
We prototype a PCB-realized tunable load network whose ports serve as additional ``virtual'' VNA ports in a ``Virtual VNA'' measurement setup. The latter enables the estimation of a many-port antenna array's scattering matrix with a few-port VNA, without any reconnections. We experimentally validate the approach for various eight-element antenna arrays in an anechoic chamber in the $700-900$~MHz regime. We also improve the noise robustness of a step of the ``Virtual VNA'' post-processing algorithms by leveraging spectral correlations. 
Altogether, our PCB-realized VNA Extension Kit offers a scalable solution to characterize very large antenna arrays because of its low cost, small footprint, fully automated operation, and modular nature.
\end{abstract}


\section{Introduction}
Antenna arrays (AAs) play a crucial role in modern wireless communications~\cite{foschini1998limits,telatar1999capacity} and sensing systems, with a clear trend toward AAs comprising ever more antenna elements. Most applications require knowledge of the AA's scattering matrix, e.g., to assess the AA's reflected power, the isolation between pairs of AA ports, or correlations between the thermal noise at different AA ports~\cite{wedge1991noise}. Both hardware specifications (e.g., for filters and amplifiers in the transmit/receive circuits) and algorithmic design (e.g., for beamforming) must be adapted to the AA's scattering matrix. 
Conventionally, the scattering matrix is measured by connecting the AA ports via monomodal transmission lines (e.g., coaxial cables) to the ports of a vector network analyzer (VNA). However, most available VNAs only have two or four ports whereas AAs with tens or hundreds of antenna elements are now envisioned and prototyped. Scaling up the number of VNA ports is cost-prohibitive. At the same time, characterizing an $N$-port AA with an $N_\mathrm{A}$-port VNA, where $N_\mathrm{A}<N$, is time-consuming and error-prone because it requires many manual reconnections between AA ports and VNA ports while terminating unconnected AA ports with matched loads~\cite{tippet1982rigorous,ruttan2008multiport,2023paper}. 

A possible solution to measure an $N$-port AA's scattering matrix with an $N_\mathrm{A}$-port VNA \textit{without any reconnections} was recently proposed with the ``Virtual VNA'' concept wherein the $N_\mathrm{S}=N-N_\mathrm{A}$ AA ports not connected to the VNA are connected to a specific $N_\mathrm{S}$-port tunable load network~\cite{del2024virtual,del2024virtual2p0,del2025virtual3p0}.\footnote{Detailed comparisons of the ``Virtual VNA'' concept with other works~\cite{garbacz1964determination,bauer1974embedding,mayhan1994technique,davidovitz1995reconstruction,wiesbeck1998wide,lu2000port,lu2003multiport,pfeiffer2005characterization,pfeiffer2005recursive,pfeiffer2005equivalent,pfeiffer2005characterization,pursula2008backscattering,bories2010small,denicke2012application,monsalve2013multiport,van2020verification,sahin2021noncontact,buck2022measuring,kruglov2023contactless,sol2024experimentally,sol2024optimal,shilinkov2024antenna,del2025physics} tackling related problems, often under simplifying assumptions, were provided in~\cite{del2024virtual,del2024virtual2p0,del2025virtual3p0} and are not repeated here for brevity. } The $N_\mathrm{S}$ ports of the tunable load network act as additional ``virtual'' VNA ports because the AA's full $N\times N$ scattering matrix can be estimated free of any ambiguities and without any assumptions beyond linearity, passivity and time-invariance. The combination of the actual VNA with $N_\mathrm{A}$ ports and the tunable load network with $N_\mathrm{S}$ ports, together with corresponding post-processing techniques, constitutes the ``Virtual VNA'' with $N=N_\mathrm{A}+N_\mathrm{S}$ ports. The requirements are $N_\mathrm{A}>1$ and the possibility to connect each port of the tunable load network to three distinct known individual loads and via a known coupled load to neighboring ports. 

Here, we report the first realization of the required tunable load network on a printed circuit board (PCB), paving the path to a low-cost, small-footprint, fully automated, and modularly scalable realization of the ``Virtual VNA'' paradigm. We experimentally test this ``VNA Extension Kit'' to characterize eight-element AAs with a four-port VNA in an anechoic chamber. Moreover, by leveraging spectral correlations, we improve the robustness of a step in the ``Virtual VNA'' algorithms for reciprocal AAs that eliminates sign ambiguities based on the coupled loads.

\textit{Organization:} In Sec.~\ref{sec_princple} we formalize the ``Virtual VNA'' concept and explain how we improve the algorithm to leverage spectral correlations. In Sec.~\ref{sec_vna_extension_kit}, we describe our VNA Extension Kit in terms of its design, measured characteristics and usage. In Sec.~\ref{sec_exp_results}, we report experimental AA characterizations based on our VNA Extension Kit. We close in Sec.~\ref{sec_conclusion} with a brief conclusion.

\section{Principle}
\label{sec_princple}

\begin{figure}[t]
    \centering
    \includegraphics[width=\columnwidth]{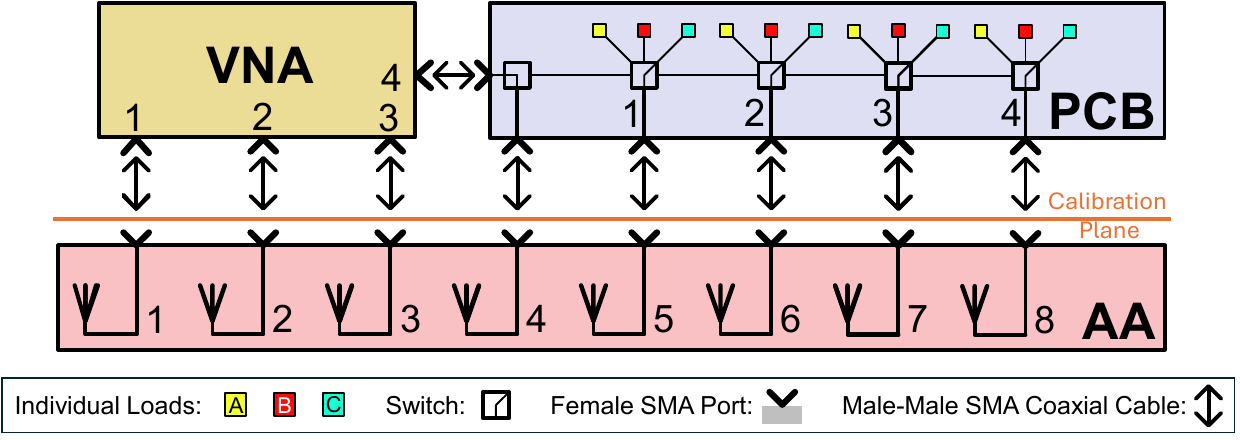}
    \caption{Schematic illustrating the ``Virtual VNA'' concept for $N_\mathrm{A}=N_\mathrm{S}=4$.}
    \label{Fig1}
\end{figure}

\textit{Problem Statement:} Our goal is to estimate the scattering matrix $\mathbf{S}\in\mathbb{C}^{N\times N}$ that characterizes a linear, passive, time-invariant AA with $N$ lumped monomodal ports. We partition the AA's ports into $N_\mathrm{A}$ ``accessible'' ports that are connected to the $N_\mathrm{A}$-port VNA and $N_\mathrm{S}$ ``not-directly-accessible'' (NDA) ports that are connected to the VNA Extension Kit. The sets $\mathcal{A}$ and $\mathcal{S}$ comprise the indices associated with the AA's accessible ports and NDA ports, respectively. As seen in Fig.~\ref{Fig1}, the VNA Extension Kit enables us to terminate the AA's $i$th NDA port with three individual loads whose reflection coefficients $r_i^\mathrm{A}$, $r_i^\mathrm{B}$ and $r_i^\mathrm{C}$ are known. $r_i^\mathrm{A}$, $r_i^\mathrm{B}$ and $r_i^\mathrm{C}$ must be mutually distinct but there is no requirement for any of them to match a calibration standard (e.g., a matched load) or for them to be identical for all NDA ports. Moreover, the kit enables us to connect the $i$th and $(i+1)$th NDA ports via a reciprocal coupled load characterized by its known scattering matrix $\mathbf{S}^{\mathrm{2PLN},i+1}\in\mathbb{C}^{2\times 2}$. Finally, the kit enables the connection of the last accessible and first NDA ports via a reciprocal coupled load characterized by $\mathbf{S}^{\mathrm{2PLN},1}\in\mathbb{C}^{2\times 2}$. 

The described hardware requirements for the VNA Extension Kit are the same for a reciprocal~\cite{del2024virtual,del2024virtual2p0} and non-reciprocal~\cite{del2025virtual3p0} AA. The role of the coupled loads is to eliminate row-wise and column-wise ambiguities which are complex-valued scaling factors in the non-reciprocal case~\cite{del2025virtual3p0} that collapse to sign ambiguities in the reciprocal case~\cite{del2024virtual,del2024virtual2p0}.

The actual VNA must have at least two ports (i.e., $N_\mathrm{A}>1$) of which the last ($N_\mathrm{A}$th) port is connected to the AA via the VNA Extension Kit; thereby it is possible to switch between connecting the last ($N_\mathrm{A}$th) accessible AA port to the VNA or to the first NDA port via a coupled load (see Fig.~\ref{Fig1}).

The $N_\mathrm{A}\times N_\mathrm{A}$ scattering matrix measured by the VNA  depends on how the AA's NDA ports are terminated by the tunable load network (which is determined by the configuration of the VNA Extension Kit). Based on measurements for a set of different configurations, the ``Virtual VNA'' post-processing algorithms recover the AA's sought-after full $N\times N$ scattering matrix without any ambiguity.

\textit{``Virtual VNA'' Algorithms:} Two families of ``Virtual VNA'' algorithms exist: The \textit{closed-form} algorithms require a specific set of configurations of the VNA Extension Kit, whereas the \textit{gradient-descent} algorithms can be applied to any random set of configurations. The details of these algorithms can be found in~\cite{del2024virtual,del2024virtual2p0,del2025virtual3p0} and are not repeated here for brevity.

Whereas existing ``Virtual VNA'' algorithms~\cite{del2024virtual,del2024virtual2p0,del2025virtual3p0} treat each frequency point independently, we here make a first step toward leveraging spectral correlations to improve the robustness of the sign ambiguity lifting in the case of reciprocal AAs. This algorithmic improvement hinges on two aspects: (i) lifting a sign ambiguity is a binary decision; (ii) the phase of any scattering coefficient must evolve continuously between adjacent frequency points provided sufficiently fine frequency sampling~\cite{yeo2022time}. 

The sign ambiguity is the same for the transmission coefficients from all accessible ports to the $i$th NDA port. First, we seek to impose the same sign ambiguity at all frequencies. To that end, we flip the sign at the $k$th frequency point $f_k$ for the row and column of the estimated scattering matrix corresponding to the $i$th NDA port if $\sum_j v_{ij}<0$ for $i\in\mathcal{S}$ and $j\in\mathcal{A}$, where
\begin{equation}
v_{i,j}=\begin{cases}
-1 & \mathrm{if\ } p_{ij}(k)>q_{ij}(k)\\
1 & \mathrm{otherwise}  
\end{cases}
\end{equation}
and
\begin{subequations}
\begin{equation}
    p_{ij}(k) = \left| \mathrm{e}^{\jmath\left[\mathrm{arg}(S_{ij}(f_k))\right]} - \mathrm{e}^{\jmath\left[\mathrm{arg}(S_{ij}(f_{k-1}))\right]} \right|,
\end{equation}
\begin{equation}
    q_{ij}(k) = \left| \mathrm{e}^{\jmath\left[\mathrm{arg}(S_{ij}(f_k))+\pi\right]} - \mathrm{e}^{\jmath\left[\mathrm{arg}(S_{ij}(f_{k-1}))\right]} \right|.
\end{equation}
\end{subequations}
Second, we seek to simultaneously lift the (now identical) sign ambiguities for all scattering coefficients associated with the $i$th NDA port at all frequency points. To that end, we flip their sign if that yields a smaller mismatch between measurement and prediction of the ``measurable'' scattering matrix at the AA's accessible ports corresponding to a configuration of the VNA Extension Kit involving the $i$th coupled load. Making this binary decision collectively for all frequency points improves its noise robustness.

\section{VNA Extension Kit}
\label{sec_vna_extension_kit}

\begin{figure}[b]
    \centering
    \includegraphics[width=\columnwidth]{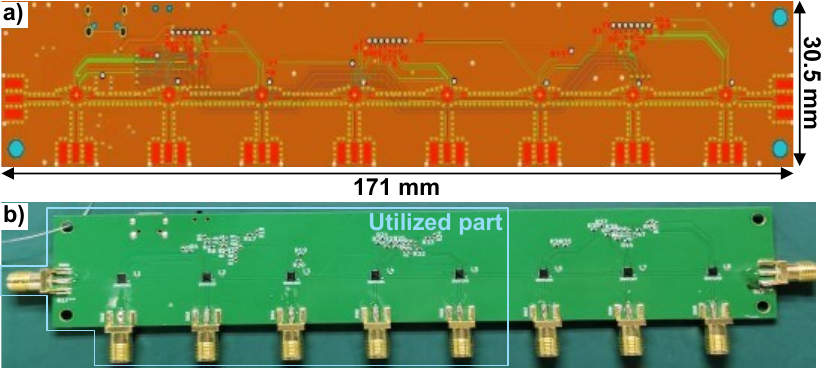}
    \caption{a) PCB layout viewed from the top layer. b) Photographic image of the manufactured PCB, highlighting the part used in the experiments below.}
    \label{Fig2}
\end{figure}

\textit{Design:} Our VNA Extension Kit is a PCB fabricated with a 4-layer FR4 structure. As seen in Fig.~\ref{Fig2}, our PCB incorporates eight SKY13414-485LF switches, each capable of connecting one of the PCB's ports to five distinct terminations. Three terminations are individual loads, realized as a 2.55~mm long stub, another stub of the same length but terminated with a 4~nH inductor, and the internal 50~$\mathrm{\Omega}$ termination of the switch. The other two terminations are connections to the two neighboring switches. 

Overall, our PCB hence offers seven ``virtual'' VNA ports (recall that the first switch is required for the AA's last accessible port). However, we only use four of these seven ``virtual'' VNA ports in our experiments below. Note also the modular nature of our PCB: A simple SubMiniature version A (SMA) connector can link two identical copies of our PCB to add eight further ``virtual'' VNA ports, etc.

The control of each switch is facilitated by 3-bit analog dc signals. Three 8-bit shift registers are employed. The reconfiguration process involves transmitting a 24-bit vector from MATLAB or Python via USB serial communication to the STM32F103C8T6 microcontroller integrated into the PCB. The STM32 then segments this vector into three 8-bit vectors and shifts the registers accordingly. 

Compared to earlier proof-of-principle demonstrations of the ``Virtual VNA'' concept, our PCB-realized VNA Extension Kit enables fully automated switching between all required loads with a small footprint and at a low cost. These aspects, combined with its modular nature, may enable scaling to very large numbers of additional ``virtual'' VNA ports.

\begin{figure}
    \centering
    \includegraphics[width=\columnwidth]{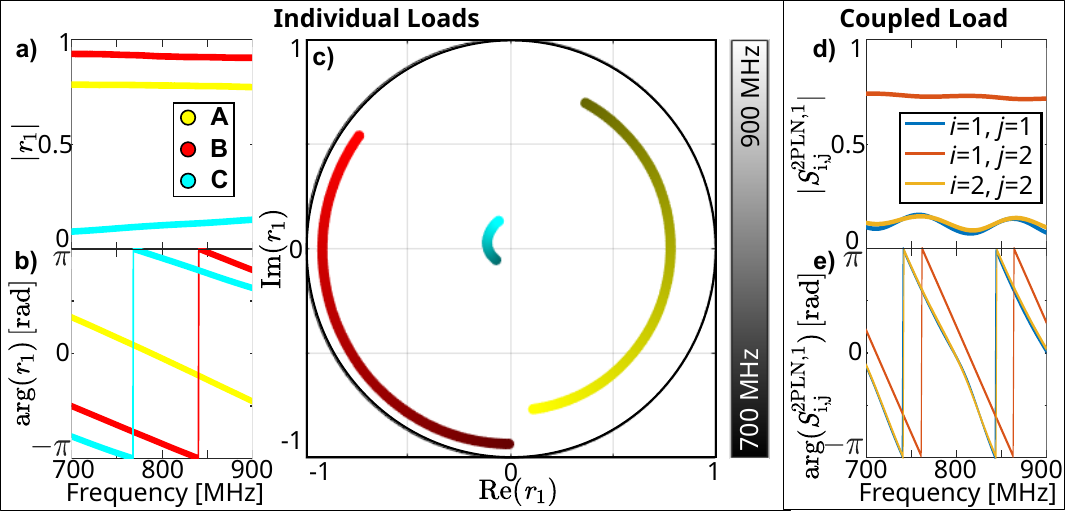}
    \caption{Measured characteristics of the PCB-realized VNA Extension Kit between 700~MHz and 900~MHz. a-c) Three individual loads in terms of their magnitudes (a), phases (b) and complex-plane representation (c). d-e) Coupled load between neighboring ports of the VNA Extension Kit in terms of magnitude (d) and phase (e) of its transmission and reflection coefficients.}
    \label{Fig3}
\end{figure}

\textit{Characterization:} Measured characteristics of our PCB-realized VNA Extension Kit are displayed for the first ``virtual'' VNA port in Fig.~\ref{Fig3} for the frequency range of interest ($700-900$~MHz). The characteristics do not substantially differ for other ``virtual'' VNA ports. None of the three individual loads emulates an ideal calibration standard (due to the propagation in the switch and microstrips) but all three individual loads are clearly distinct from each other at all frequencies, which is the ``Virtual VNA'' algorithms' only requirement for the individual loads. Meanwhile, the coupled load (which is simply a reciprocal microstrip transmission line) displays a strong transmission coefficient, in line with the requirement that its transmission coefficients should not vanish (or else the coupled load would not serve its purpose).

\textit{Usage:} Similar to a VNA that needs to be calibrated before usage, our VNA Extension Kit needs to be calibrated before usage as part of a ``Virtual VNA''. Specifically, the reflection coefficients of the individual loads and the scattering matrices of the coupled loads must be measured. These calibration measurements require a two-port VNA (which is compatible with our requirement $N_\mathrm{A}>1$ for the ``Virtual VNA''). The transmission lines that will be used to connect the VNA Extension Kit to the AA are treated as part of the VNA Extension Kit, as seen from the calibration plane in Fig.~\ref{Fig1}.\footnote{A technical subtlety is that the gender of the VNA Extension Kit's connectors is the opposite of that of the AA's connectors. Hence, the VNA calibration used to characterize the VNA Extension Kit is different from the one that will be used in the main measurements even if $N_\mathrm{A}=2$.}
Once the VNA Extension Kit is calibrated, the $N_\mathrm{A}$-port VNA used for the main measurements must be calibrated. The transmission lines that will be used to connect the VNA to the AA are treated as part of the VNA, as seen from the calibration plane in Fig.~\ref{Fig1}.\footnote{The transmission line from the last (in our case: fourth) VNA port to the corresponding AA port is routed via the VNA Extension Kit, as shown in Fig.~\ref{Fig1}, and accordingly included in the VNA calibration.}

Having calibrated the $N_\mathrm{S}$-port VNA Extension Kit and the $N_\mathrm{A}$-port VNA, we connect the $N$-port AA to the $N$-port ``Virtual VNA'' comprising the actual VNA and the VNA Extension Kit. Then, we measure the $N_\mathrm{A}\times N_\mathrm{A}$ scattering matrix with the actual VNA for different configurations of the VNA Extension Kit in order to subsequently estimate the full $N \times N$ scattering matrix of the AA, as mentioned earlier.

\section{Experimental AA Characterization \\via ``Virtual VNA'' Measurements }
\label{sec_exp_results}

\begin{figure*}
    \centering
    \includegraphics[width=\textwidth]{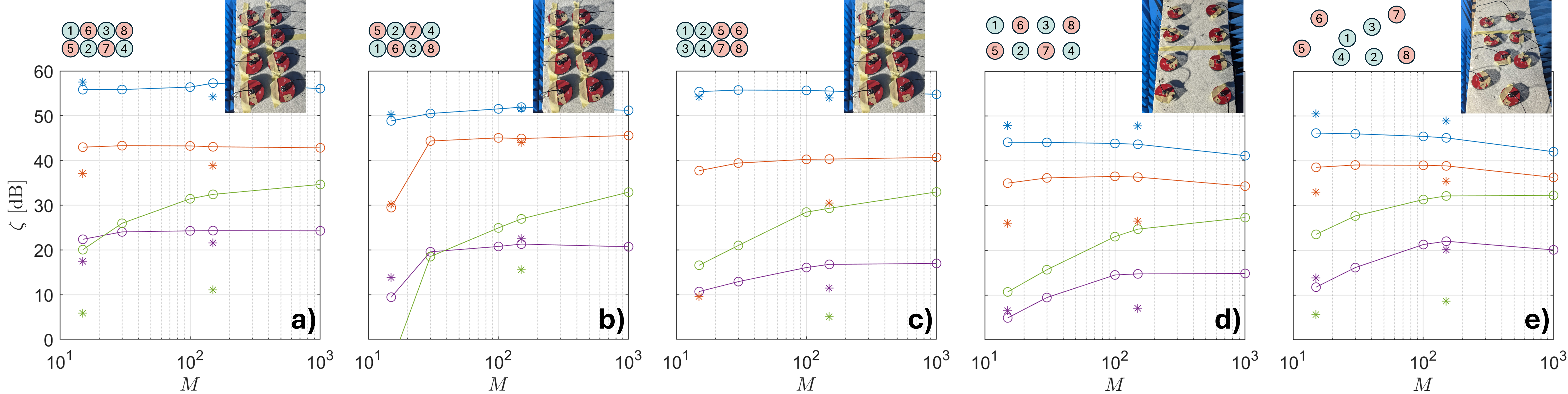}
    \caption{Influence of the number $M$ of measurements with individual load configurations on the accuracy metric $\zeta$ evaluated for different groups of scattering coefficients (blue: $\mathcal{AA}$ block; red: $\mathcal{AS}$ block; purple: $\mathcal{SS}$ block diagonal; green: $\mathcal{SS}$ block off-diagonal) for the two considered algorithms (star: closed form; circle: gradient descent) for five distinct setups (depicted above subplots).}
    \label{Fig5}
\end{figure*}

We experimentally test our PCB-based VNA Extension Kit as part of a ``Virtual VNA'' to characterize three eight-element reciprocal AAs (antenna element type: AEACBK081014-S698) in an anechoic chamber. As seen in Fig.~\ref{Fig1}, four AA ports ($\mathcal{A}=\{1,2,3,4\}$) are connected to a Keysight P5024B four-port VNA (IFBW: 500~Hz; power: 13 dBm; 201 frequency points); the other four AA ports ($\mathcal{S}=\{5,6,7,8\}$) are connected to ``virtual'' VNA ports of our VNA Extension Kit. As seen in Fig.~\ref{Fig5}, the first AA is a $2\times 4$ regular array with 8~cm center-to-center spacing for which we consider three different port indexing choices; the second AA is a $2\times 4$ regular array with 20~cm center-to-center spacing; the third AA is a random array. For each of these five setups, we measure ten times the data required for the closed-form algorithm (to be able to average) as well as the data required for the gradient-descent algorithm (with 1000 random realizations of individual load configurations) -- see details in Fig.~2E in~\cite{del2024virtual2p0}. We run both types of ``Virtual VNA'' post-processing algorithms for different values of the number $M$ of measurements with individual load configurations. In each case, we quantify the accuracy of the estimated AA scattering matrix for different groups of scattering coefficients via the following metric that is defined similar to a signal-to-noise ratio, treating the error as ``noise'':
\begin{equation}
    \zeta = \left\langle\frac{\mathrm{SD}\left[S_{ij}^\mathrm{GT}(f)\right]}{\mathrm{SD}\left[S_{ij}^\mathrm{GT}(f) - S_{ij}^\mathrm{PRED}(f)\right]}\right\rangle_{i,j,f},
    \label{eq_zeta}
\end{equation}
where $\mathrm{SD}$ denotes the standard deviation, and the superscripts GT and PRED denote ground truth and prediction, respectively. We measure the ground truth with two cascaded Keysight P5024B four-port VNAs (acting like one eight-port VNA).

\begin{figure}
    \centering
    \includegraphics[width=\columnwidth]{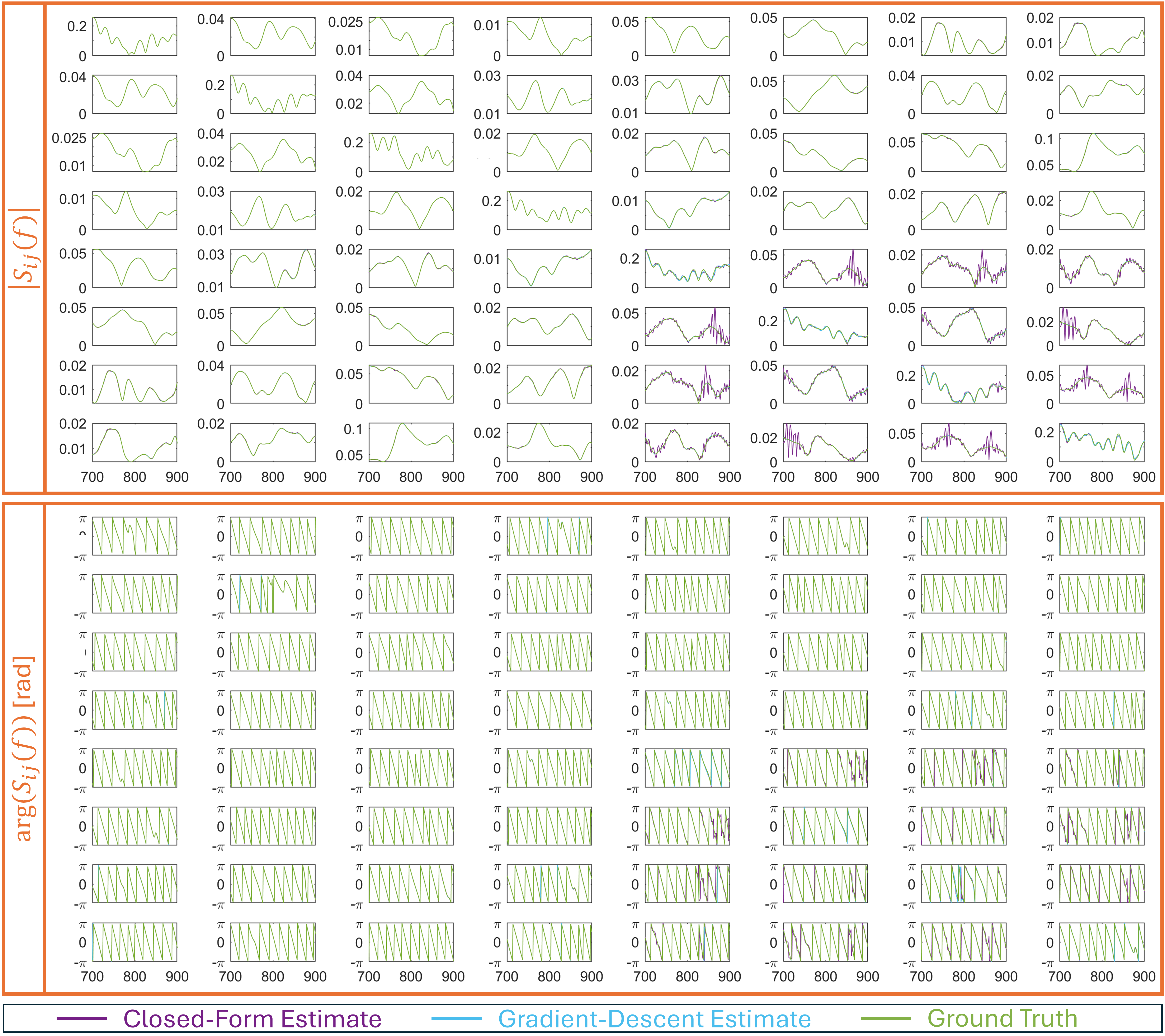}
    \caption{Superposed display of estimated (closed form [purple] and gradient descent [cyan]) and ground-truth [green] scattering spectra for Setup 1 in terms of magnitudes [top] and phases [bottom]. The ground-truth spectra are printed on top; when only the green line is visible, the estimates are flawless.}
    \label{Fig4}
\end{figure}

To start, we visually compare in Fig.~\ref{Fig4} for Setup 1 the ground-truth scattering spectra and our two estimates thereof (based on averages over the ten repeated measurements for the closed-form algorithm; based on all 1000 random configuration measurements for the gradient-descent algorithm). Both estimates are flawless in the $\mathcal{AA}$, $\mathcal{AS}$ and $\mathcal{SA}$ blocks. The gradient-descent algorithm also yields faithful estimates in the $\mathcal{SS}$ block in which the closed-form algorithm notably struggles with off-diagonal entries, especially in terms of their magnitudes. Because the ``Virtual VNA'' approach leverages minute changes of the ``measurable'' scattering matrix at the AA's accessible ports as a result of changes in the VNA Extension Kit's configuration, the weaker the coupling between the AA ports is, the more vulnerability to measurement noise there is. The current experiments in an anechoic chamber are for this reason more challenging than previous experiments in a reverberation chamber in~\cite{del2024virtual,del2024virtual2p0}. We attribute the superiority of the gradient-descent algorithm regarding the $\mathcal{SS}$ block to the fact that the changes of the measurable scattering matrix between subsequent measurements are larger (because more than one or two ``virtual'' VNA port terminations are changed), and that more distinct VNA Extension Kit configurations are used.

The systematic quantitative analysis in terms of $\zeta$ in Fig.~\ref{Fig5} reveals additional insights. First, as expected, the accuracy is largest for the most tightly coupled AA. Second, the port indexing matters: the results for Setup 1 are comparable in Fig.~\ref{Fig5}(a) and Fig.~\ref{Fig5}(b) but slightly inferior in Fig.~\ref{Fig5}(c) where NDA ports were spatially clustered. Third, whereas the gradient-descent algorithm achieves higher accuracies for off-diagonal than diagonal entries of the $\mathcal{SS}$ block, the opposite applies to the closed-form algorithm. In the gradient-descent algorithm, only the estimation of the off-diagonal entries of the $\mathcal{SS}$ block appears to notably benefit from very large values of $M$. The decline in accuracy for all other entries for very large values of $M$ seen in Fig.~\ref{Fig5}(d,e) is attributed to thermal drift in the experiment (the last measurements for the gradient-descent algorithm are taken hours after the ground-truth measurement).

\section{Conclusion}
\label{sec_conclusion}

To summarize, we have prototyped and experimentally tested a PCB-based VNA Extension Kit which realizes the specific tunable load network required by the ``Virtual VNA'' concept. We successfully estimated the scattering matrices of various eight-element AAs with a four-port VNA \textit{without any reconnections} because the ports of this tunable load network serve as additional ``virtual'' VNA ports. Along the way, we improved the ``Virtual VNA'' algorithms by leveraging spectral correlations to make the lifting of sign ambiguities more noise robust. Thanks to its small footprint, low cost, full automation, and modular nature, our PCB-based VNA Extension Kit may enable a scalable solution to characterizing very large AAs.

Looking forward, on the hardware side, we will work on designing an ultra-broadband-deployable version of our VNA Extension Kit. Meanwhile, on the software side, we will work on identifying additional means of leveraging spectral correlations to further improve the accuracy.
Incidentally, when all AA ports are connected to our ``VNA Extension Kit'', the resulting system can be interpreted as a ``beyond-diagonal reconfigurable intelligent surface'' (BD-RIS)~\cite{shen2021modeling}, specifically, a tri-diagonal RIS.

\bibliographystyle{IEEEtran}

\providecommand{\noopsort}[1]{}\providecommand{\singleletter}[1]{#1}%

\end{document}